# Curriculum Design of Competitive Programming: a Contest-based Approach

Zhongtang Luo

───────────── ✦ ─────────────

**Abstract**—Competitive programming (CP) has been increasingly integrated into computer science curricula worldwide due to its efficacy in enhancing students' algorithmic reasoning and problem-solving skills. However, existing CP curriculum designs predominantly employ a problem-based approach, lacking the critical dimension of time pressure of real competitive programming contests. Such constraints are prevalent not only in programming contests but also in various real-world scenarios, including technical interviews, software development sprints, and hackathons.

To bridge this gap, we introduce a contest-based approach to curriculum design that explicitly incorporates realistic contest scenarios into formative assessments, simulating authentic competitive programming experiences. This paper details the design and implementation of such a course at Purdue University, structured to systematically develop students' observational skills, algorithmic techniques, and efficient coding and debugging practices. We outline a pedagogical framework comprising cooperative learning strategies, contest-based assessments, and supplemental activities to boost students' problem-solving capabilities.

## 1 INTRODUCTION

Competitive programming (CP), the practice of solving and programming pre-defined algorithm problems, has been widely recognized as an effective way to improve students' problem-solving skills and algorithmic thinking. On one hand, theoretical analysis has shown that CP problems highly align with computer science curriculum guidelines [1]. On the other hand, empirical evidences have also demonstrated that CP has considerable educational benefit, including measurable improvement in students' problem-solving skills, motivation, as well as retention rates [2], [3]. As a result, educators around the world have started to incorporate elements of CP into their courses, including Brazil [2], [3], Mainland China [4], [5], Hong Kong [6], India [7], and the United States [8].

However, as we analyze these literatures, most curriculum designs of competitive programming are *problem-based*: a set of algorithmic problems are assigned with very loose due-date constraints, and students can choose to solve them at their own pace. While such designs are not without their own merit, we observe that these courses left out the time factor in competitive programming: in a real contest, participants are required to solve problems under demanding time constraints. For example, in the ICPC contest series [9], teams are usually given 10 or more problems to solve in 5 hours with one computer; in Codeforces [10] — one of the most popular online programming contest platforms — participants are given 2 hours to solve 5 problems. All of these translate to less than 30 minutes of computer time per problem. Therefore, these contests not only test students' algorithmic knowledge, but also their ability to code efficiently under pressure.

TABLE 1
Comparison between the three courses.

| Course | Focus |
|---|---|
| CS 21100 | Basic observation skills. Some techniques on graph problems. |
| CS 31100 | Summarization and expansion of CP1 observation skills. Basic techniques on various topics. |
| CS 41100 | Review of CP1 observation skills and CP2 techniques. Advanced techniques and implementation practices. |

We argue that such time pressure is not artificial but present in many real-world scenarios as well. For example, the Scrum framework [11] requires developers to produce results daily and teams to complete a 'sprint' every 1–4 weeks. Technical interviews are known to be stressful, and part of the stress is also attributed to the time factor [12]. Various activities, such as game jams hackathons, also require participants to produce results in a very limited time. Therefore, we believe that time pressure is ubiquitous in computer science in the real world.

Therefore, to promote better performance in programming contests and real-world scenarios, we propose a *contest-based* curriculum design of competitive programming, where formative assessments are designed to mimic the time pressure of real contests. In this paper, we will present our course design at Purdue University and discuss various trade-offs.

## 2 ORGANIZATION OF THE COURSE

In Purdue University, competitive programming courses are divided into three levels: CS 21100, CS 31100 and CS 41100, known as CP1, CP2 and CP3, respectively. Each course is a 2-credit 12-week short course, and builds on the prior course to expand the students' knowledge of competitive programming. A comparison of the focuses of the three courses can be found in Table 1.

This curriculum design is focused on CS 41100, although we believe the ideas can be applied to other levels as well.

# 3 COURSE DESIGN

In this section, we present our course design based on discussion of content, assessment and pedagogy.

## 3.1 Content

The overarching idea of our course is to enable students to solve algorithmic problems efficiently. Unfortunately, problem-solving has long been known to involve extensive tacit knowledge [13]. Therefore, we aim to define our enduring outcomes in a way that sheds dome light on the outcomes we deem necessary for problem-solving under the computer science context.

### 3.1.1 Enduring Outcomes

Based on our anecdotal experience, we consider that competitive programming problems need three aspects of skill: *observation*, *technique*, and *implementation* [14]. Loosely speaking, observation is the observational ability to reduce an unknown problem to an easier problem; technique represents the already known knowledge about algorithms and data structures; and implementation is the ability to translate the solution into a working program. We note that while there are no clear boundaries between these three aspects, some topics may still be considered to be more observational, as a greater emphasis is placed on understanding and reducing the problem, such as greedy, dynamic programming, and combinatorial problems. Other topics may be more technical, requiring more fixed knowledge, such as geometry, segment trees, and strings. Therefore, we articulate our enduring outcomes based on these three aspects:

EO-1. **Observation skills** reduce a new algorithmic problem to a known problem that can be solved.

EO-2. **Techniques** solve known algorithmic problems efficiently.

EO-3. **Implementation by coding and debugging** builds a solution.

### 3.1.2 Important-to-Know & Supplemental Outcomes

Given that the topic of the course is competitive programming, we aim for student agency in our important-to-know outcomes. Specifically, since the field of competitive programming is vast, and there is simply no way that any course can cover all the topics, we show our students ways to continue learning and practicing after the course to further their knowledge of the enduring outcomes. We define our important-to-know outcomes as such:

IO-1. Algorithm design is an iterative process that involves trial and error and building upon previous iterations.

IO-2. Relevant techniques to solve known problems can be researched and studied online.

IO-3. Efficiency in algorithm design, coding, and debugging can be gained through repeated practice.

For supplemental outcomes, we focus on ICPC — the most relevant contests for college students. Given the special format of ICPC contests requiring extensive teamwork and reference, we define our supplemental outcomes accordingly:

SO-1. To succeed in competitive programming, one must allocate time strategically and cooperate with teammates.

SO-2. Special techniques and templates are prepared and used to help competitive programming.

SO-3. Setting competitive programming problems requires creativity and a deep understanding of relevant concepts.

# 4 ASSESSMENT

In this course, we will try to give formative assessments that resembles competitive programming contests. We recognize that similar to timed tests, contests often introduce test anxiety if students have a reason to worry about their performance [15]. Therefore, we place the bulk of our grading on post-contest activities, while provide some incentive for students to perform well in the contests. In this way, we hope to provide a safe net for students to alleviate their anxiety, while simultaneously providing a realistic environment for them to practice and adapt to the time pressure.

## 4.1 Learning Objectives

We base our assessment on the following learning objectives:

Students will be able to ...

LO-1. Select and apply appropriate techniques covered in CP1 (*DFS/BFS, Shortest Path, Floodfill, Topological Sort, Tarjan, Union Find Set, Minimum Spanning Tree*), CP2 (*Binary Exponentiation, Linear Sieve, ExGCD, Combinatorics, Inclusion-Exclusion, Sparse Table, Fenwick Tree, Basic Computational Geometry, Convex Hull, Rolling Hash, Trie*) and CP3 (*Half-Plane Intersection, Adaptive Simpson, Linear Recurrence, Segment Tree, LCA, HLD, Network Flow, KMP, AC Automata*) to solve known algorithmic problems. (EO-2)

LO-2. Implement and debug a solution to algorithmic problems efficiently. (EO-3)

LO-3. Combine LO-1 – LO-3 to solve algorithmic problems, performing at Candidate Master (rating 1900+) level in Codeforces contests. (EO-1 – EO-3, IO-1 – IO-3)

## 4.2 Performance Task — Contests

When designing our contest, we need to budget it under the logistic limit of our class time. While we believe that an ICPC style contest of five hours provides immense practice for our students and helps boost their experience with real contests, we recognize that such time slot is not feasible in a classroom setting. In the end, we decide to adopt a 90-minute contest which fits in most classroom settings with minimal adjustment.

In line with the three enduring outcomes, we decide to provide three problems every contest session. Given that contest participants usually pick and solve problems within their reach rather than solving every of them, we emphasize to our students that they are not expected to solve every problem in the contest. Instead, we encourage them to bring unsolved problems back home and upsolve them as homework before the due date. We provide a detailed rule set for the contest in Section A.1.1 in the sample syllabus.

We also provide a sample contest with three problems in Section C in the appendices.

To design our grading rubric, we consider the expected performance of students in our class. A perfect student, in our opinion, should be able to solve 2 problems every session. On the lower end, we believe that students who demonstrate an acceptable understanding of our course should be able to upsolve every problem after the contest. We budget in standard leniency of two free drops and three late days to account for difficulties and unexpected events. We provide a detailed grade scheme in Section A.2 in the sample syllabus.

### 4.3 Supplemental Activities

During the evolution of this course over the years, we have considered various ways to allow the students to further their understanding of competitive programming. We list a few of our attempts below.

**Note Sharing** In the first iteration of the course, we have designed a note sharing program that allows interested students to take notes for the class and submit them for extra credit. A total of 6 points can be earned through this program, with the following rubric:

1) 2 points for the general flow of the course and concepts;
2) 2 points for the first problem discussed in the class;
3) 2 points for the second problem discussed in the class.

We find that while the students are motivated to sign up and provide notes, the notes often become a verbatim reiteration of what happened during the lecture with numerous mistakes, and are very costly both for the instructor and the student to fix. As a result, we provide detailed slides in the next iteration of the course and discontinue the program.

**Student-Proposed Topics** In order to encourage students to explore and study topics that are not covered in the course (IO-2), we provide a list of advanced topics for students to vote in class and study individually. We also allow students to team up and present the topic for extra credit. We set the following requirement for the extra credit:

1) Study the topic and solve the sample problem.
2) Create a reference that helps with problem-solving of the topic.
3) Present the topic, the solution and the reference to the class.
4) Propose and solve 3 suitable problems to be used for that week's contest.

A rubric matching the requirements is provided in Table 2.

To accommodate for students who are not able to do a presentation, we allow them to submit a self-study report on the topic they selected for extra credits. We grade the report on:

1) Provide a short tutorial to the topic. It should be understandable to the average student in CP3. Wrong and/or unintelligible explanation will reduce the point. (2 pts)
2) Pick one problem that requires the technique. It should not be solvable by something significantly easier than the topic. Provide a tutorial on how to solve the problem. It should be understandable to the average student in CP3. Wrong and/or unintelligible explanation will reduce the point. (2 pts)

3) Implement and solve the problem. (2 pts)

Overall, we find that the students are engaged with the student presentation and provide much discussion. However, we also find that the presentation has a tendency to focus on the implementation details of the topic rather than insights of the general problem-solving observations. For instance, in the fast Fourier transform, students may be more inclined to discuss operations of complex number and the butterfly diagram in great details, while leaving out the part of how to form the problem as polynomials that can be multiplied together.

We think that the project is a good way to provide student motivation and engagement, but presentation quality may be sacrificed a bit as the student (rather than the instructor) is doing the work. We did not include the project in the next iteration of the course, but we may bring it back in the future.

**Problem Setting** In the most recent iteration of the course, we have introduced a problem setting project that allows students to propose and set a competitive programming problem for future uses. This is in line with our supplemental outcomes SO-3. We provide a detailed description in Section A.1.1 in the sample syllabus.

A total of 3 problems have been set by the students. We find that the students are able to provide interesting problems with good difficulty. However, we also find that given that the problems are new and not tested, the instructor needs to spend a significant amount of time reviewing the problems and providing feedback.

## 5 PEDAGOGY

We observe that a course in competitive programming is significantly different from a course in algorithms and complexity. In this course, we often care little about the theoretic underpinning of the algorithms, but rather how to reduce unknown problems and solve them efficiently. This section reflects our understanding of the pedagogy of this course.

### 5.1 Course Schedule

We think the course schedule can be somewhat arbitrary, since in competitive programming, the topics are like a web that does not show a clear hierarchy. However, we consider the following rationales when designing the course schedule:

1) We start with implementation and geometry problems. This is to introduce students to implementation details (EO-3), which is often overlooked in lower levels of competitive programming.
2) We start the topic of range queries, including Fenwick trees, RMQ and segment trees, early. This is because range queries are common tools in advanced competitive programming, and this arrangement allows us to interlace various applications of range queries in different topics throughout the course.
3) We separate topics by being more observation-heavy or technique-heavy. We distribute them throughout the course, and set up a team contest for technique-heavy topics. This is because technique-heavy topics tend to demand more coding, and we believe a team effort can reduce the workload for individual students.

TABLE 2
Rubric for presentation of student-proposed topics.

| Criteria | A (2pts/ea.) | B (1.5pts/ea.) | C (1pts/ea.) | D (0pts/ea.) |
|---|---|---|---|---|
| Study | Solve the sample problem and prepare a reference. | Solve the sample problem. | Propose a solution to the sample problem. | Not able to solve the problem. |
| Presentation | Present the technique, the problem, the solution, and the reference. Explain how the technique and the solution work. Go through the code and the reference in detail. | Present the technique, the problem, and the solution. Explain how the technique and the solution work. Go through the code in detail. | Present the technique and the problem to the class. Explain how the technique works. | Not able to present anything. |
| Proposal | Propose and solve 3 suitable problems. | Propose and solve 2 suitable problems. | Propose and solve 1 suitable problem. | Not able to propose any suitable problem. |

If chosen, the student automatically gets 2 points for every problem he/she proposes in the contest. The student also gets an extra maximum of 6 points, subject to the rubric above.

The expectation is that every student should provide 2-3 codes (for a total of 6) and do either the recitation on the topic or 1-2 problems. In case of significant under-performing by some student, the effort will be graded separately.

A tentative schedule is provided in Section A.2.2 in the sample syllabus.

## 5.2 Lesson Plan

The plan for one lecture involves three activities:
1) code presentation for the previous contest; (30 min)
2) introduction of new concepts; (15 min, or as needed)
3) in-class problem discussion. (15–30 min per problem)

**Code Presentation** In this activity, we invite the first solver for every problem of the previous contest to introduce their solution and present their code. We believe that this activity is beneficial for multiple reasons.

1) It provides a good review for the previous week. One common issue of competitive programming classes is that since the topics are often scattered, it is easy to forget what happened in the course. This activity reviews the material after one week and aims to improve retention.
2) It aids student engagement and motivation. It is believed that having students describe their challenges in the class and how they overcome them can help motivate other students [16]. When students describe how they solve difficult problems, they set a role model that inspires other students as well.
3) It helps with implementation practice. Efficient implementation (EO-3) is a skill-based knowledge that is often difficult to acquire. By having students present their code, we can discuss various implementation details and help students improve their coding skills based on the example.

**Introduction of New Concepts** When introducing new concepts, we believe that instead of providing a detailed description of the theoretical underpinning of the concept, it is more beneficial to provide an interface of the concept that can be used in problem-solving, say "see the reference for a piece of code, and google it if you want to learn more theory," and then dive into a problem that requires the concept. For example, when teaching network flow, it is enough for the students to understand the network flow problem and that there are algorithms in the reference that solve the problem. Due to the time limit of the course, it is impractical and unnecessary to introduce detailed algorithms such as Dinic or Edmonds-Karp, since understanding these algorithms takes a lot of time yet contributes little to the actual problem-solving.

**In-Class Problem Discussion** We spend the bulk of the class time on in-class problem discussion. We believe that since observation skill (EO-1) is tacit knowledge that can only be demonstrated on specific problems, it is important to provide students with a variety of problems to learn and practice the skill.

We give a handout every class that contains the learning objectives and the problems. We select these problems from real contests to provide an authentic experience for the students. We provide a sample handout in Section B in the appendices.

Given the relative small size of the class, it is feasible for the instructor to assign a problem, and then walk around to engage with the students about their thinking. We also observe that students simultaneously discuss their thoughts with each other as the problems are often at a higher difficulty level.

## 5.3 Cooperative Learning

Given that competitive programming requires extensive tacit skills, we believe group-based cooperative learning can help students to learn from each other [17].

To achieve this goal, we designed team contests that require students to work in teams of two or three, mimicking the ICPC contest format (see Section A.1.1 in the sample syllabus for detailed rules). Given that there are vast skill gaps between different students, we use the result from the previous individual contest as a criterion to assign one-third of the students as team leaders. We then distribute the rest of the students randomly.

To facilitate cooperation, we give full points to every student whose team solves a problem. This provides extrinsic motivation for students to cooperate and help each other. To avoid the free rider problem, we require that each team member may solve at most one problem in the contest. In practice, we observe that the team often allocates easier problems to the less experienced team members, and vise-versa. We consider it a good outcome that every student in the team is working on a problem that is at their level.

Given that the problems in the contest are often at a high difficulty level, we observe that students tend to simultaneously engage in detailed discussion with team members during the contest, and report that team contests feel easier than individual contests. We also observe that the team leaders often take the role of explaining the solution to the team members and overseeing their coding process. We believe that these interactions are beneficial for the students to learn from each other.

## 5.4 Use of Code Reference

Similar to the ICPC contest and online contests, we allow students to bring unlimited printed materials to the contest (see Section A.1.1 in the sample syllabus). We ban the use of copy-pasting code from the computer to avoid students storing solutions for every problem. We believe that the use of code reference is beneficial because the students can focus on problem-solving instead of memorizing algorithm codes. We also provide a code reference that contains everything the class needs on GitHub [18].

## 5.5 Difficult Concepts

During the course, we observe that students often struggle with some tacit concepts that competitive programming differs from other computer science subjects. We list a few of them below.

1) **Negative transfer from other algorithm courses.** Competitive programming drastically differs from other algorithm courses in that it forgoes rigorous theoretical analysis and focuses on problem-solving.

   a) **Intuition is more important than proof.** In competitive programming, having an idea of something that may work is often enough. Anecdotally, we find that fixiation on the proof can harm one's observation skills (EO-1) if they are afraid of coding or reasoning based on a guessed hunch. In contrast, it may be helpful to emphasize to students that they can guess anything reasonable as long as they cannot find a counterexample [19], and success in competitive programming is defined by having a solution that works for the test cases.

   b) **Implementation matters!** Students from algorithm courses often find implementation-heavy problems (EO-3) to be boring as they don't seem to carry any theoretical weight. However, we observe that ICPC contests often have implementation-heavy problems that must be addressed. For example, every ICPC World Finals has a geometry problem that requires extensive implementation. In ICPC ECNA 2023, four problems in the less solved half are geometry/pure search problems. We find that it is often helpful to emphasize that it is worth considering the easiest way to implement a solution. For instance, in grid searching, the following code

```
int dx[] = {1, 0, -1, 0};
int dy[] = {0, 1, 0, -1};
for (int i = 0; i < 4; i++) {
    int nx = x + dx[i], ny = y + dy[i];
    // ...
}
```

is often easier to write and debug than repeating the same logic four times. We also find that emphasizing to students that software engineering jobs often require extensive implementation skills can help motivate them.

2) **Don't force a technique onto a problem.** We find that sometimes knowing the topic of the week can create an adverse effect to the students, as they tend to treat the technique as a panacea ritual and stop thinking altogether. The effect is most pronounced in general techniques such as segment trees, as students sometimes draw a segment tree and try to blindly feed input data into the tree, rather than thinking about how to solve the problem first. We find this story [20] resonates with students pretty well:

   *Knowing what to do with numbers is certainly the heart and soul of basic arithmetic. However, knowing what to do can mean rather different things. I have always been charmed by this example from an elementary school child reported a number of years ago:*

   *I know what to do by looking at the examples. If there are only two numbers I subtract. If there are lots of numbers I add. If there are just two numbers and one is smaller than the other it is a hard problem. I divide to see if it comes out even and if it doesn't I multiply.*

   During the in-class problem discussion, it may be beneficial to explicitly show the difference between solving a problem and then applying a segment tree at the last step and drawing a segment tree and then trying to blindly fit data in.

# 6 EXECUTION OF THE COURSE

We taught the class twice in Spring 2024 and Spring 2025. The students may take the class either for normal grades or for pass/fail. Of the 14 students that took the course in 2024, 6 chose normal grades and 8 chose pass/fail. All the 6 students attained an A or above grade, and all the 8 students attained a P grade.

Incidentally, students in the course also represented Purdue University in the ICPC 2022 and 2024 series. The 2022 team advanced to the World Finals, and the 2024 team advanced to the North America Championship.

# 7 FUTURE IMPROVEMENTS

No design is perfect. We list a few of the improvements we can make in the future.

1) **Build more scaffolding for topics.** We observe that there exists a significant gap from acquiring knowledge in class to immediately applying it in the contest. Therefore, it may be beneficial to provide one scaffolding problem together with the tutorial that may be solved before the contest. The problem will be graded based on completion and can serve as a gentle warm-up for the contest.

2) **Extra credit on improving the reference.** We find that students often underappreciate the importance of code

reference (SO-2), sometimes even showing up to the contest without one. As a result, we can provide extra credit for students who help to improve the current reference. The student will propose the scope of the improvement and the corresponding credit. Upon approval and completion, the student will receive the credit. We hope that this activity helps the students to build a sense of ownership of the reference and improve their problem-solving skills through self-guided exploration (IO-2).

## 8 Conclusion

In this paper, we proposed a contest-based curriculum design for competitive programming (CP), aimed at addressing the limitations of existing problem-based approaches by incorporating realistic time pressures into formative assessments. By embedding authentic contest scenarios into the educational framework, our curriculum explicitly emphasizes observational skills, technical knowledge, and efficient implementation — three critical competencies required for success in both competitive programming contests and real-world computing scenarios.

Initial results from deploying our contest-based approach show promising outcomes, with students achieving significant academic success and demonstrating competitive excellence in programming contests. Ultimately, our curriculum aims not only to produce competent competitive programmers but also skilled, agile problem-solvers capable of navigating complex challenges in diverse computing contexts.

# APPENDIX A
# SAMPLE SYLLABUS

We provide a sample syllabus for the course.

## A.1 Course Description

CP3 teaches experienced programmers additional techniques to solve competitive programming problems and builds on material learned in CP1 and CP2. This includes algorithmic techniques. Primarily, CP3 prepares students to compete in programming contests, which means most class time is focused on simulating contest environments and teaching teamwork and communication alongside problem practice.

The course revolves around three aspects that are essential in problem-solving. Together they form the enduring outcomes of the course:

EO-1. **Observation skills** reduce a new algorithmic problem to a known problem that can be solved.

EO-2. **Techniques** solve known algorithmic problems efficiently.

EO-3. **Implementation** by coding and debugging builds a solution.

A comparison of CP1, CP2 and CP3 is available below:

| Course | Focus |
|---|---|
| CS 21100 | Basic observation skills. Some techniques on graph problems. |
| CS 31100 | Summarization and expansion of CP1 observation skills. Basic techniques on various topics. |
| CS 41100 | Review of CP1 observation skills and CP2 techniques. Advanced techniques and implementation practices. |

### A.1.1 Learning Outcomes & Assessment

Students will be able to ...

LO-1. Select and use appropriate observation skills covered in CP1 (*Search, Greedy, Dynamic Programming, BSTA*), CP2 (*Pruning, Perspective, Sweep Line, Monotonic Queue, Dynamic Programming on Tree, DFS Order on Tree, Bitmask*) and CP3 (*Monotonicity, Offline*) to understand and reduce problems to known algorithmic problems. (EO-1)

LO-2. Select and apply appropriate techniques covered in CP1 (*DFS/BFS, Shortest Path, Floodfill, Topological Sort, Tarjan, Union Find Set, Minimum Spanning Tree*), CP2 (*Binary Exponentiation, Linear Sieve, ExGCD, Combinatorics, Inclusion-Exclusion, Sparse Table, Fenwick Tree, Basic Computational Geometry, Convex Hull, Rolling Hash, Trie*) and CP3 (*Half-Plane Intersection, Adaptive Simpson, Linear Recurrence, Segment Tree, LCA, HLD, Network Flow, KMP, AC Automata*) to solve known algorithmic problems. (EO-2)

LO-3. Implement and debug a solution to algorithmic problems efficiently. (EO-3)

LO-4. Combine LO-1 – LO-3 to solve algorithmic problems, performing at Candidate Master (rating 1900+) level in Codeforces contests. (EO-1–EO-3)

**Performance Task** To assess LO-1–LO-4, the student is provided with three competitive programming problems every assessment session (contest). The online judge website grades every problem on a pass/fail basis. The student gets instant feedback and can revise within the time frame. The student is either assessed individually or in teams of three. The student is tasked to solve the problems with the following rules, consistent with ICPC-style contests:

1) You have 90 minutes to solve 3 problems in class. Problems solved in class net 2 points.

2) However, you are not required or expected to solve every problem in class! You may bring unsolved problems back home and upsolve them as homework before the due date. Every problem nets 1 point as homework.

3) As a rule of thumb, you need 4 points every week for an A, and 2 points every week for a P over 12 sessions. 2 lowest performances are dropped when calculating the grade.

4) You may bring your laptop or use the lab computer. However, cellphones and the Internet are **not** allowed other than submission.

5) Instead, you may bring a printed-out reference to help you solve the problem. You must type the solution yourself.

6) For educational purposes, we may ask you to present your code in class.

7) To track your progress in the course, you are required to fill in the solve information and your reflections in a Google Sheets document. We will provide the relevant template in class.

8) **For team contest:** One teammate is allowed to code **no more than** one problem in class. You should discuss and decide who codes what after reading the problems. A solve in the team counts as 2 points for every team member. (For teams of two, you may attempt the third problem with anyone as long as the first two solves are done by different people. To put it more elegantly and generally, the difference of solves between team members in contest should not be greater than one.)

9) **For team contest:** Homeworks (upsolving) are still individual and no plagiarism is allowed. Only the coder can use the partial solution code he/she writes in class if there is one.

Problem Selection For the three problems every session:

1) Problem 1 reviews and examines the student's knowledge from CP1 and CP2. The problem's solution involves skills and techniques from these two prior classes, with a focus on the observation skill (EO-1).

2) Problem 2 assesses the student's understanding of the current topic in CP3. The problem's solution involves techniques from the latest topic (EO-2).

3) Problem 3 assesses the student's implementation abilities. The problem involves extensive coding and debugging and requires a mastery of these abilities from the student (EO-3).

The order of the problems may be randomized.

Upsolving Trying to solve these problems during the contest session is not all there is to this course! In fact, what you successfully do in the contest is no more than a practice and review of topics you already understand. On the other hand, what you failed to solve in the contest is a golden opportunity to compare yourself with other students, identify your own weaknesses and do some catch-up. **For this reason, I ask you to write a very short reflection (around** 10 words) **on every problem in the Google Sheets document, together with the solve information.** The following checklist may be helpful for you to reflect on your progress on every problem after every contest.

Level 1. **I solved the problem within the allotted time.**

Congratulations! You have demonstrated your capabilities in the contest. If you feel that there is still room to improve your efficiency, you are more than welcome to utilize the advice below. Otherwise, feel free to move on.

Level 2. **I solved the problem after the contest.**

A perfect student should be able to get 2 problems every session. If you feel you are ever stuck for a significant period in the contest for some reason, feel free to check below on the reason you get stuck for some advice. You can also drop in during my office hours and we can walk through your contest process to identify the ways to improve your efficiency.

Level 3. **I know how to solve the problem, but my code runs into some bugs.**

Debugging is often a painful process for many people, including myself. You may want to try a variety of debugging methods (i.e. static, dynamic, stress test, etc.). You may find it helpful to write down the exact bug (integer overflow, typo, logic issues, etc.) in the reflection, as knowing what bug happens most frequently in your code helps you to speed up your debugging process significantly.

You can also drop in during any office hours if you cannot find the bug in your code.

Level 4. **I know how to solve the problem, but I don't know how to implement it.**

Unfortunately, as we enter CP3, the complexity in the code's logic steadily increases, and you will start to face what we call implementation-heavy problems. These problems resemble what you will face in your future software engineering jobs: not so algorithmically complicated, but rather a bunch of logic that you have to implement correctly. I often find it helpful to take out a pen, think about and write down the structure of the code. What is the most concise way to implement the logic? How many functions do I need to make? What is the logic flow in each function? As I think through these aspects, I gradually begin to grasp what I need to code and become more comfortable to code it.

Level 5. **I know the related technique, but I don't know how to solve this specific problem.**

As we encounter more and more techniques in CP2 and CP3, a common difficulty is to link the technique to the problems we are solving. In general, I do not find it very helpful to get stuck thinking about any problem for more than 1 hour if you have absolutely no idea how to solve it. Instead, after we go over the solution in the lecture, try to ask yourself: *How could I have thought the solution myself? What is the observation I am missing here?* In other words, you need to identify a possible way for you to come up with the solution yourself so that you do not get stuck in future contests on a similar problem.

If you find it difficult to conjure a possible way for you to come up with the solution yourself, feel free to talk to me after class or during office hours.

Level 6. **I don't know the related technique.**

In a sense this is the easiest thing to fix: you already understand what you don't know! Fortunately, there are a lot of online resources for every technique we covered in CP1, CP2 and CP3. I generally found Google to be reasonably helpful. You may need to glance over a couple of different blogs before you have a more comprehensive understanding, but that is totally fine. You can also talk to me after class or during office hours and I can either explain or help you find relevant resources.

**Problem Setting (Extra Credit)** Students may earn up to 6 extra points by proposing and setting a competitive programming problem for future uses.

- Students may work in teams of at most 3. Everyone in the team is expected to have a similar workload. Notably, **everyone is expected to code a solution** to ensure accuracy.
- Students should use Polygon (https://polygon.code forces.com) to develop the problem. The problem is expected to be complete, with a statement, a checker, a validator, tests, solutions (2+ AC and 1+ TLE, if applicable), and a tutorial. Students should create a problem and add the instructor to review the problem.
- Students should submit the draft of a statement by **Topic 3**, the draft of a complete problem by **Topic 6**, and a complete problem by **Topic 9**. The instructor will review the drafts and provide feedback. For this reason, late submissions may not be accepted.
- The problem might be used in the course or school contests. For this reason we ask that you **do not discuss the problem with other people**.
- The problem will be graded by the rubric in Table 3.

## A.2 Grade

The following scale is used to assign a grade over 10 assessment sessions, **after dropping the 2 lowest performances from a total of 12 sessions**:

| Grade | A+ | A | A- | B+ | B | B- | C+ | C |
|-------|-----|-----|-----|-----|-----|-----|-----|-----|
| Point | 45 | 40 | 38 | 35 | 30 | 28 | 25 | 20 |
| | C- (P) | D+ (NP) | D | D- | F | | | |
| | 18 | 15 | 10 | 8 | ≤ 7 | | | |

**Late Policy** Every student is given three free late days in total for upsolves to account for unexpected events. For extraordinary circumstances, please contact the instructor. In addition, 2 lowest performance sessions are dropped when calculating the grade.

**Attendance** Since for contests, an unannounced absence creates an unfair burden on the rest of the team and unnecessary stress on the logistics of the course, students absent from contest sessions without being excused by the instructor will be given a **0** for that contest with **no** opportunities for upsolve. For this course, **arriving more than 10 minutes** late or **leaving more than 10 minutes early** without being excused counts as missing the class. In the event of an anticipated absence, inform the instructor of the situation as far in advance as possible.

TABLE 3
Rubric for problem setting.

| Grade | 2 pts | 1 pts | 0 pts |
|---|---|---|---|
| Difficulty | The problem is of appropriate or higher difficulty for CP3. | The problem is of appropriate difficulty for CP1 or CP2. | The problem is trivial. |
| Completeness | The problem is complete. | The problem lacks enough solutions to ensure accuracy or lacks a tutorial. | The problem is not complete. |
| Clarity | The statement and tutorial is clear and easy to follow. | The statement or tutorial needs clarification. | The statement or tutorial cannot be understood. |

How to Succeed in this Course Below are a few tips that might help you.

- **Prepare for the stress.** Solving problems in a strictly-timed environment is a stressful event for everyone. Unfortunately, this is a common occurrence in competitive programming and interview exams. In general, having more contest experience helps, so this course employs multiple low-stake contest sessions with free drops available to minimize stress. You may also find participating in online contests on Codeforces and At-Coder helps to build a rich competition experience, as they provide some incentive (rating) but are also generally low-stakes.
- **Be an active thinker in the contest.** As you approach CP3 and beyond, you will start to find that implementation starts to become easier but coming up with ideas for the problems remains difficult. To practice solving the problems, you need to be efficient not only in implementing a solution but also in coming up with ideas, and the best way to practice is to think about the contest problems in class actively instead of relying on your teammates.
- **Study and survey a wide range of sources for one topic.** For most topics in this class (and in competitive programming in general), there are many different resources online. The ability to search and study these resources is essential to your success in competitive programming and problem-solving, even outside of this class. In general, do not be afraid to glance over 10 articles online and find the one that helps you understand the topic more thoroughly. You will need this skill once you graduate from the course and study yourself.

### A.2.1 Logistics

The following resources and platforms are used in this course.

**Helpful Resources** None of the following resources are required for the course. Nevertheless, they may help you understand competitive programming and topics in the class.

- *Competitive Programming 4* by Halim, Halim, and Effendy
- *Algorithms for Competitive Programming*: https://cp-algorithms.com/index.html

- *Codeforces Catalog*: https://codeforces.com/catalog

**Brightspace** Announcements and grades are published on Brightspace. Please check regularly for announcements that you may have missed.

**Codeforces** Assessment sessions will be done on Codeforces (https://codeforces.com/).

- You need to create an account on the platform if you do not have one already and provide us with the account name. You also need to join the Codeforces group for this class. Sign-up details will be announced in class.
- For team-based contests, you need to create a team (https://codeforces.com/teams) with your teammates added to the team before you register for the contest.

**Google Sheets** We use Google Sheets to help you track your progress throughout the course. You are required to fill in your progress and your team's progress on Google Sheets. Details will be announced in class.

### A.2.2 Tentative Course Schedule

| Topic (Tuesday) | Contest (Thursday) |
|---|---|
| Topic 0: Introduction, Implementation | Individual Contest |
| Topic 1: Geometry: Review, Half-plane Intersection, Adaptive Simpson | Team Contest |
| Topic 2: Combinatorics: Review, Linear Recurrence | Individual Contest |
| Topic 3: Range Query: Review, Segment Tree Hard | Team Contest |
| Topic 4: Monotonicity: DP Review, Optimization | Individual Contest |
| Topic 5: Tree: Review, LCA, HLD | Team Contest |
| Topic 6: Game Theory: SG Function, Search | Individual Contest |
| Topic 7: Network Flow: Min Cut, Min Cost | Team Contest |
| Topic 8: Fast Fourier Transform | Individual Contest |
| Topic 9: String: KMP, AC Automata | Team Contest |
| Topic 10: Offline: CDQ, Mo's Algorithm | Individual Contest |
| Topic 11: Final Contest | Team Contest |

# Appendix B
## Sample Handout

We provide a sample handout for *Topic 7: Network Flow: Min Cut, Min Cost*.

### Learning Objectives

The students will be able to...
1) describe the **max-flow network flow problem** and the **min-cost network flow problem**;
2) describe the **max-flow min-cut theorem**;
3) apply well-established algorithms (**ISAP**, **Dinic**, **EK**, **ZKW**, etc.) to solve a network flow problem;
4) model programming problems (e.g. a **matching problem**) with network flow.

### Sample Problems

**Problem Name:** Magic Potion
**Link:** https://vjudge.net/problem/Gym-101981I

**Problem Name:** Kejin Game
**Link:** https://vjudge.net/problem/UVALive-7264

**Problem Name:** Coding Contest
**Link:** https://vjudge.net/problem/HDU-5988

**Problem Name:** Hiring Employees
**Link:** https://dmoj.ca/problem/noi08p3

### Magic Potion

There are $n$ heroes and $m$ monsters living on an island. The monsters have become very vicious recently, so the heroes have decided to reduce their numbers. However, the $i$-th hero can only kill one monster from the set $M_i$. Joe, the strategist, has $k$ bottles of magic potion, each of which can buff one hero's power, allowing him to kill one additional monster. Since the potion is very powerful, a hero can only take at most one bottle of potion.

Please help Joe determine the maximum number of monsters that can be killed by the heroes if he uses the optimal strategy.

#### Input

The first line contains three integers $n$, $m$, $k$ ($1 \le n, m, k \le 500$) — the number of heroes, the number of monsters, and the number of bottles of potion.

Each of the next $n$ lines contains one integer $t_i$, the size of $M_i$, followed by $t_i$ integers $M_{i,j}$ ($1 \le j \le t_i$), the indices (1-based) of monsters that can be killed by the $i$-th hero ($1 \le t_i \le m$, $1 \le M_{i,j} \le m$).

#### Output

Print the maximum number of monsters that can be killed by the heroes.

#### Examples

**Input**
```
3 5 2
4 1 2 3 5
2 2 5
2 1 2
```
**Output**
```
4
```
**Input**
```
5 10 2
2 3 10
5 1 3 4 6 10
5 3 4 6 8 9
3 1 9 10
5 1 3 6 7 10
```
**Output**
```
7
```

#### Source

2018 ACM-ICPC Asia Nanjing Regional Programming Contest

## Kejin Game

In recent years, many free-to-play games, referred to as Kejin games, have emerged. These games are accessible without charge, but specific items or characters require payment. Examples include Love Live, Kankore, Puzzle & Dragon, Touken Ranbu, and Kakusansei Million Arthur, all of which have gained immense popularity and generate significant revenue daily.

In a Kejin game, your character possesses a skill graph that determines how skills can be acquired. This graph is a directed acyclic graph where vertices represent skills, and edges indicate dependencies: if there is an edge from skill A to skill B, A is a prerequisite for B. In cases where skill S has multiple dependencies, all must be acquired before obtaining S. Furthermore, each edge in the graph is unique, and cyclic dependencies are absent.

Acquiring skills typically involves time and effort, especially for advanced skills deeper in the graph. However, as a player with resources to spare, you can choose to pay money, denoted as "Ke," to bypass certain restrictions. Specifically, money can be used to remove edges from the dependency graph or to acquire skills directly, ignoring existing dependencies.

Given the constraints of time and money, you wish to optimize the balance between them. Each action, be it acquiring a skill through conventional means, removing an edge, or directly purchasing a skill, incurs a cost measured in units of "TA." You seek to determine the minimal cost required to acquire a desired skill S, starting without any initial skills.

### Input

The input starts with an integer indicating the number of test cases (no more than 10). Each test case consists of:

- The first line containing three integers $N$, $M$, and $S$, where $N$ is the number of vertices (skills), $M$ is the number of arcs (dependencies), and $S$ is the index of the target skill (1-based index).
- $M$ subsequent lines each contain three integers $A$, $B$, and $C$, where there is an arc from skill $A$ to skill $B$ with $C$ TAs cost to remove the dependency.
- A line with $N$ integers representing the cost to acquire each skill through normal means.
- Another line with $N$ integers representing the cost to directly acquire each skill via payment, bypassing any dependencies.

The costs for acquiring skills or removing dependencies range up to $1,000,000$.

### Output

For each test case, output a single line containing the minimal cost to acquire the specified skill $S$.

### Examples

**Input**

```
2
5 5 5
1 2 5
1 3 5
2 4 8
4 5 10
3 5 15
3 5 7 9 11
100 100 100 200 200
5 5 5
1 2 5
1 3 5
2 4 8
4 5 10
3 5 15
3 5 7 9 11
5 5 5 50 50
```

**Output**

```
31
26
```

*Source*

2015 ACM-ICPC Asia Beijing Regional Contest

## Coding Contest

A coding contest will be held in this university, in a huge playground. The whole playground will be divided into $N$ blocks, and there will be $M$ directed paths linking these blocks. The $i$-th path goes from the $u_i$-th block to the $v_i$-th block. Your task is to solve the lunch issue. According to the arrangement, there are $s_i$ competitors in the $i$-th block. Limited to the size of table, $b_i$ bags of lunch including breads, sausages, and milk would be put in the $i$-th block. As a result, some competitors need to move to another block to access lunch.

However, the playground is temporary, and as a result there would be many wires on the path. For the $i$-th path, the wires have been stabilized at first and the first competitor who walks through it would not break the wires. Since then, however, when a person goes through the $i$-th path, there is a chance of $p_i$ to touch the wires and affect the whole network. Moreover, to protect these wires, no more than $c_i$ competitors are allowed to walk through the $i$-th path. Now you need to find a way for all competitors to get their lunch, and minimize the possibility of network crashing.

### Input

The first line of input contains an integer $t$ which is the number of test cases. Then $t$ test cases follow. For each test case, the first line consists of two integers $N$ ($N \leq 100$) and $M$ ($M \leq 5000$). Each of the next $N$ lines contains two integers $s_i$ and $b_i$ ($s_i, b_i \leq 200$). Each of the next $M$ lines contains three integers $u_i$, $v_i$, and $c_i$ ($c_i \leq 100$) and a floating-point number $p_i$ ($0 < p_i < 1$). It is guaranteed that there is at least one way to let every competitor have lunch.

### Output

For each test case, output the minimum possibility that the network would break down. Round it to 2 digits.

### Examples

**Input**

```
1
4 4
2 0
0 3
3 0
0 3
1 2 5 0.5
3 2 5 0.5
1 4 5 0.5
3 4 5 0.5
```

**Output**

```
0.50
```

### Source

2016 ACM-ICPC Asia Qingdao Regional Contest

## Hiring Employees

BuBu has stepped into a challenging role as the head of the human resources department for a subsidiary of the Olympic committee. His task is to recruit a team of employees for a new Olympic project. The project spans $N$ days, with day $i$ requiring at least $A_i$ employees.

The company has $M$ types of employees available for hire. Employees of type $i$ work from day $S_i$ to day $T_i$ and require a total salary of $C_i$ dollars. BuBu's goal is to minimize the total cost of hiring enough employees for all necessary days.

### Input

The first line of input contains two integers $N$ and $M$, where $N$ is the number of days and $M$ is the number of employee types. The second line has $N$ nonnegative integers, representing the minimum number of employees required each day. The next $M$ lines each contain three integers $S_i$, $T_i$, and $C_i$, describing the availability and cost of each type of employee.

For 100% of the test cases, $1 \leq N \leq 1000$, and $1 \leq M \leq 10000$. Also, other values in the data will not exceed $2^{31} - 1$.

### Output

Output one integer, the cost of the optimal hiring strategy.

### Examples

**Input**

```
3 3
2 3 4
1 2 2
2 3 5
3 3 2
```

**Output**

```
14
```

### Source

2008 China National Olympiad in Informatics

# Appendix C
# Sample Contest

We provide a sample contest for *Topic 7: Network Flow: Min Cut, Min Cost*.

## Parencedence!

Parencedence is a brand new two-player game that is sweeping the country (that country happens to be Liechtenstein, but no matter). The game is played as follows: a computer produces an arithmetic expression made up of integer values and the binary operators +, −, and *. There are no parentheses in the expression. If Player 1 goes first, he/she can put parentheses around any one operator and its two operands; the parenthesized expression is evaluated, and its value is used in its place. Player 2 then does the same, and the game proceeds accordingly, Player 1 and Player 2 alternating turns. Player 1's object is to maximize the final value, while Player 2's object is to minimize it. A sample round might go as follows:

- Initial expression: $3 − 6 * 4 − 7 + 12$
- Player 1's move: $3 − 6 * (4 − 7) + 12 \rightarrow 3 − 6 * −3 + 12$
- Player 2's move: $(3 − 6) * −3 + 12 \rightarrow −3 * −3 + 12$
- Player 1's move: $(−3 * −3) + 12 \rightarrow 9 + 12$
- Player 2's move: $(9 + 12) \rightarrow 21$

A game of Parencedence is played in two rounds, each using the same initial unparenthesized expression: in the first round, Player 1 goes first, and in the second, Player 2 goes first (Player 1 is always trying to maximize the result and Player 2 is always trying to minimize the result in both rounds, regardless of who goes first). Let $r_1$ be the result of the first round and $r_2$ the result of the second round. If $r_1 > −r_2$, then Player 1 wins; if $r_1 < −r_2$, then Player 2 wins; otherwise the game ends in a tie. Your job is to write a program to determine the final result, assuming both players play as well as possible.

### Input

The first line of the input file contains an integer $n$ indicating the number of test cases. The test cases follow, one per line, each consisting of a positive integer $m \leq 9$ followed by an arithmetic expression. The value of $m$ indicates the number of binary operators in the arithmetic expression. The only operators used will be +, −, and *. The − operator can appear as both a unary and a binary operator. All binary operators will be surrounded by a single space on each side. There will be no space after any unary −. No combination of parentheses will ever result in an integer overflow or underflow.

### Output

For each test case, output the case number followed by three lines. The first contains the first set of operands and operator to be parenthesized in round 1 (when Player 1 goes first) and $r_1$. The second line contains the analogous output for round 2. The third line contains either the phrase "Player 1 wins", "Player 2 wins" or "Tie" depending on the values of $r_1$ and $r_2$. In the first two output lines, if there is a choice between which operator should be parenthesized first, use the one which comes earliest in the original expression. Follow the format used in the examples.

### Examples

**Input**

```
2
4 3 − 6 * 4 − 7 + 12
2 45 − −67 − 3
```

## Output

```
Case 1:
Player 1 (7+12) leads to -2
Player 2 (3-6) leads to -27
Player 2 wins
Case 2:
Player 1 (-67-3) leads to 115
Player 2 (45--67) leads to 109
Player 1 wins
```

### Source

## Stampede!

You have an $n \times n$ game board. Some squares contain obstacles, except the left- and right-most columns which are obstacle-free. The left-most column is filled with your $n$ pieces, 1 per row. Your goal is to move all your pieces to the right-most column as quickly as possible. In a given turn, you can move each piece N, S, E, or W one space, or leave that piece in place. A piece cannot move onto a square containing an obstacle, nor may two pieces move to the same square on the same turn. All pieces move simultaneously, so one may move to a location currently occupied by another piece so long as that piece itself moves elsewhere at the same time.

Given $n$ and the obstacles, determine the fewest number of turns needed to get all your pieces to the right-hand side of the board.

### Input

Each test case starts with a positive integer $n$ indicating the size of the game board, with $n \leq 25$. Following this will be $n$ lines containing $n$ characters each. If the $j^{th}$ character in the $i^{th}$ line is an 'X', then there is an obstacle in board location $i, j$; otherwise this character will be a '.' indicating no obstacle. There will never be an obstacle in the $0^{th}$ or $(n-1)^{st}$ column and there will always be at least one obstacle-free path between these two columns. A line containing a single 0 will terminate input.

### Output

For each test case output the minimum number of turns to move all the pieces from the left side of the board to the right side.

### Examples

**Input**

```
5
.....
.X...
...X.
..X..
.....
5
.X...
.X...
.X...
.XXX.
.....
0
```

**Output**

```
Case 1: 6
Case 2: 8
```

### Source

## Machine Programming

One remarkable day company "X" received $k$ machines. And they were not simple machines, they were mechanical programmers! This was the last unsuccessful step before switching to android programmers, but that's another story.

The company has now $n$ tasks, for each of them we know the start time of its execution $s_i$, the duration of its execution $t_i$, and the company profit from its completion $c_i$. Any machine can perform any task, exactly one at a time. If a machine has started to perform the task, it is busy at all moments of time from $s_i$ to $s_i + t_i - 1$, inclusive, and it cannot switch to another task.

You are required to select a set of tasks which can be done with these $k$ machines, and which will bring the maximum total profit.

### Input

The first line contains two integer numbers $n$ and $k$ ($1 \le n \le 1000$, $1 \le k \le 50$) — the numbers of tasks and machines, correspondingly.

The next $n$ lines contain space-separated groups of three integers $s_i, t_i, c_i$ ($1 \le s_i, t_i \le 10^9$, $1 \le c_i \le 10^6$), $s_i$ is the time when they start executing the $i$-th task, $t_i$ is the duration of the $i$-th task, and $c_i$ is the profit of its execution.

### Output

Print $n$ integers $x_1, x_2, \ldots, x_n$. Number $x_i$ should equal 1, if task $i$ should be completed, and otherwise it should equal 0.

If there are several optimal solutions, print any of them.

### Examples

**Input**
```
3 1
2 7 5
1 3 3
4 1 3
```
**Output**
```
0 1 1
```
**Input**
```
5 2
1 5 4
1 4 5
1 3 2
4 1 2
5 6 1
```
**Output**
```
1 1 0 0 1
```

### Source

VK Cup 2012 Round 3